\documentclass[pdflatex,sn-basic]{sn-jnl}

\usepackage{amsmath,amssymb}
\usepackage{bm}

\jyear{2021}%

\theoremstyle{thmstyleone}%
\theoremstyle{thmstyletwo}%

\theoremstyle{thmstylethree}%

\providecommand{\DIFdel}[1]{} 

\begin{document}

\title[ ]{Time-series image denoising of pressure-sensitive paint data by projected multivariate singular spectrum analysis}


\author*[1]{\fnm{Yuya} \sur{Ohmichi}}\email{ohmichi.yuya@jaxa.jp}

\author[2]{\fnm{Kohmi} \sur{Takahashi}}

\author[1]{\fnm{Kazuyuki} \sur{Nakakita}}

\affil[1]{\orgdiv{Aviation Technology Directorate}, \orgname{Japan Aerospace Exploration Agency}, \orgaddress{\street{Jindaiji-Higashi}, \city{Chofu}, \postcode{182-8522}, \state{Tokyo}, \country{Japan}}}

\affil[2]{
\orgname{AdvanceSoft Corporation}, \orgaddress{\street{Kanda-Surugadai}, \city{Chiyoda}, \postcode{101-0062}, \state{Tokyo}, \country{Japan}}}



\abstract{Time-series data, such as unsteady pressure-sensitive paint (PSP) measurement data, may contain a significant amount of random noise.
Thus, in this study, we investigated  a noise-reduction method that combines multivariate singular spectrum analysis (MSSA) with low-dimensional data representation.
MSSA is a state-space reconstruction technique that utilizes time-delay embedding, and the low-dimensional representation is achieved by projecting data onto the singular value decomposition (SVD) basis.
The noise-reduction performance of the proposed method for unsteady PSP data, i.e., the projected MSSA, is compared with that of the truncated SVD method, one of the most employed noise-reduction methods.
The result shows that the projected MSSA exhibits better performance in reducing random noise than the truncated SVD method.
Additionally, in contrast to that of the truncated SVD method, the performance of the projected MSSA is less sensitive to the truncation rank.
The projected MSSA achieves denoising effectively by extracting smooth trajectories in a state space from noisy input data.
Expectedly, the projected MSSA will be effective for reducing random noise in not only PSP measurement data, but also various high-dimensional time-series data.}

\maketitle
{\centering
Graphical Abstract
\includegraphics[width = .8\textwidth]{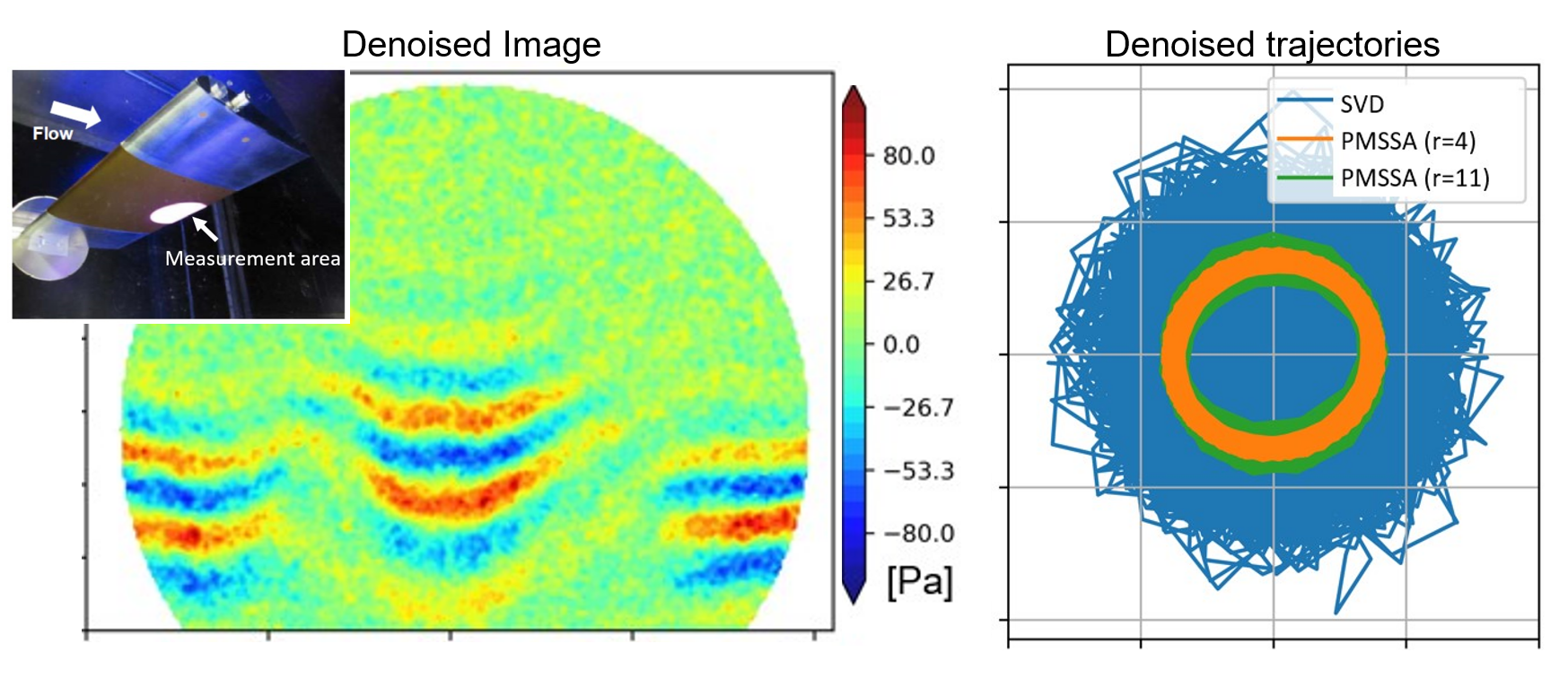}
\par
}
\section{Introduction}
Understanding the nature of physical phenomena via high-dimensional time-series images is a fundamental approach employed in scientific research.
In the field of fluid dynamics, the spatiotemporal distribution of pressure and velocity fields in a flow field can be measured using methods such as pressure-sensitive paint (PSP) \citep{Liu2021} and particle-image velocimetry \citep{Westerweel2013}, for example.
However, since experimental data usually contain unwanted signals, i.e., noise, it is necessary to properly separate the noise from the signals representing the physical phenomena.

The PSP measurement, which is the subject of this study, is an optical measurement technique that has been applied to a variety of flow fields, from low to hypersonic speeds \citep{Nakakita2011, Peng2019, Running2019}, because of its ability to measure pressure with high temporal and spatial resolution.
However, when the pressure fluctuation of the observation target is small, the PSP image is significantly affected by shot noise \citep{Jiao2019, Varigonda2021, Liu2022}. 
Shot noise is noise that is independent of each pixel, and it generates random statistical fluctuations that follow the Poisson distribution.
Generally, increasing the light intensity is one method for reducing shot noise.
However, it may be impossible to obtain signals with sufficient intensity from the unsteady PSP measurement (called fast PSP) because the exposure time is insufficient.
A common technique for reducing shot noise is spatial averaging \citep{Panda2017, Tang2021}.
However, this technique does not sufficiently reduce the noise in some cases, and it reduces the spatial resolution.

Recently, noise-reduction methods based on data-analysis techniques, such as modal decomposition and sparse sensing, have been explored.
The singular value decomposition (SVD), also called proper orthogonal decomposition (POD) \citep{Lumley1967, Lumley1993} or principal component analysis (PCA), are often used to PSP images to extract coherent patterns or reduce noise \citep{Pastuhoff2013, Gordeyev2014, Peng2016, Sugioka2019, Zhou2022}.
The SVD is employed with a reduced-rank approximation (truncated SVD) and noise is removed by reconstructing the PSP images using only modes that represent signal components.
One of the difficulties with the truncated SVD is extracting only the signal components, which is achieved by setting the appropriate truncation rank.
The choice of the appropriate truncation rank is nontrivial, and several criteria based on the cumulative sum of singular values \citep{Pastuhoff2013}, the signal periodicity \citep{Peng2016}, or reference pressure signals \citep{Sugioka2019} have been used.
Recently, \cite{Gavish2014} proposed the optimal truncation rank based on asymptotic analysis.
However, the appropriate truncation rank is problem-dependent, and it is difficult to determine the truncation rank optimally based on a single criterion.
In addition, these methods do not modify the SVD temporal coefficients that represent the temporal behavior of the modes, even though the coefficients are also affected by noise.
Modifying the temporal coefficients is expected to achieve further noise reduction and suppress the truncation rank dependence of noise-reduction performance.
\cite{Wen2018} proposed a data fusion algorithm that optimizes the temporal coefficients using a limited number of scattered clean data.
\cite{Noda2018} employed the coherent output power between PSP and sound-level meter signals to extract pressure-distribution patterns relevant to the target periodic acoustic phenomena.
These methods can make effective use of the temporal behavior information of target phenomena to reduce noise but cannot be applied to PSP data alone.
Recently, \cite{Inoue2021} proposed a method that combines SVD with sparse sensing that can modify the SVD temporal coefficients using PSP data alone.

Dynamic mode decomposition (DMD) has also been actively studied as a data-analysis method similar to SVD \citep{Schmid2010, Taira2017}.
With the DMD method, one can simultaneously obtain spatial and temporal patterns by assuming a linear time evolution.
\cite{Ali2016} applied DMD to PSP images and extracted pressure-fluctuation patterns as DMD modes.
To apply the DMD to noisy data, several algorithms that reduce the effect of noise have been proposed \citep{Hemati2017, Nonomura2019, Ohmichi2022}.
\cite{Ohmichi2022} applied the DMD algorithm for noisy data to fast PSP images of transonic flow over a swept wing \citep{Ohmichi2018, Sugioka2021}, and successfully extracted the pressure distribution related to transonic buffet phenomenon.

In fluid dynamics research, the time evolution of SVD temporal coefficients is often analyzed to understand the nature of various phenomena. 
The phase plot of the coefficients, which represents the relationship between different SVD modes, shows a smooth trajectory \citep{Sieber2016, Daniel2018}.
These researches inspired us to explore a noise-processing method that combines the low-dimensional representation of data on modal bases with multivariate singular spectrum analysis (MSSA), which is a multivariate extension of basic SSA \citep{Broomhead1986, Hassani2010}.
Basic SSA is employed for the state-space reconstruction of time-series signals using time-delay embedding, which is often employed for noise processing \citep{Golyandina2010}.
By combining MSSA and data representation using the temporal coefficients and extracting smooth trajectories in the state space that represents the temporal behavior of physical phenomena, we expect to achieve effective noise reduction for high-dimensional time-series data.
Notably, a similar approach has been proposed by \cite{Chhatkuli2015} in medical-imaging research.
They conducted a time-series forecasting for medical imaging using the MSSA combined with PCA coefficients, which are identical to the temporal coefficients of SVD.
In this study, we utilized a similar approach to reduce random noise in high-dimensional time-series data.

The objective of this study is to clarify the effectiveness of a noise-reduction method that combines MSSA with low-dimensional data representation via projection onto the SVD basis.
For brevity, this method is referred to as the projected MSSA (PMSSA) in this paper.
The projected MSSA is applied to two test cases.
One is the pressure field data of the K\'{a}rm\'{a}n vortex street behind a square cylinder obtained by a numerical fluid simulation, and the other is the practical unsteady PSP measurement data relevant to a tone trailing-edge (TE) noise of a two-dimensional airfoil.
The noise-reduction performance of the projected MSSA is compared with that of the truncated SVD (TSVD), which is one of the most common noise-reduction methods.
Additionally, the effects of the truncation rank used in the projected MSSA and the truncated SVD on the noise-reduction performance are discussed.

\section{Approach}
\subsection{Input dataset}
The input dataset comprises high-dimensional time-series data, such as time-series spatial-distribution data.
We assume that each instantaneous data, denoted as a snapshot in this paper, has $d$ elements ($d$ is the number of pixels of the images) and $j$-th snapshot $\bm{x}(j)$ can be represented as $\bm{x}(j)=[ x_{1,j}, \cdots , x_{d,j} ]^\mathsf{T} $, where $j = 1,\cdots,m$ and $m$ is the number of snapshots.
Thus, the input dataset can be represented as the matrix $X$ whose columns are $\bm{x}(j)$,

\begin{equation}
{X} =
		\begin{bmatrix}
		   \bm{x}(1) & \cdots & \bm{x}(m) 
		\end{bmatrix}
		=
\begin{bmatrix}
   x_{1,1} & \cdots  & x_{1,m}\\
   \vdots  & \ddots & \vdots\\
   x_{d,1} & \cdots  & x_{d,m}
\end{bmatrix}.
\end{equation}
That is, $X$ is a $d \times m$ matrix, where the row and column indices correspond to the spatial and temporal distributions, respectively.

\subsection{Noise reduction by truncated SVD} \label{Sec:SVD}
The input data matrix $X$ can be factorized by SVD to
\begin{equation}
    X = U S V^{\mathsf{T}} = \sum_{i=1}^{r_X}  \bm{u}_i  \sigma_i  \bm{v}_i^{\mathsf{T}}, \label{Eq:TSVD1}
\end{equation}
where $U$ and $V$ are the orthogonal matrices, and $S$ is the diagonal matrix containing the singular values.
The left and right singular vectors, $\bm u_i$ and $\bm v_i$, are the $i$-th columns of $U$ and $V$, respectively; $\sigma_i$ is the $i$-th singular value, and $r_X$ is the rank of $X$.
The singular values are arranged in descending order.
The $\bm u_i$ and $\bm v_i$ represent a spatial mode and its temporal behavior, respectively. 
If the input dataset $X$ is extremely large, hindering the application of the batch SVD algorithm, online algorithms \citep{Ross2008, Halko2011,Ohmichi2017c} can be applied.

In the noise-reduction method by the truncated SVD, we assume that the noise can be separated as the component corresponding to small singular values ($\sigma_i < \sigma_r$).
\begin{equation}
    X =  \sum_{i=1}^{r}  \bm{u}_i  \sigma_i  \bm{v}_i^{\mathsf{T}} + \sum_{i=r+1}^{r_X}  \bm{u}_i  \sigma_i  \bm{v}_i^{\mathsf{T}}. \label{Eq:TSVD2}
\end{equation}
The first and second terms in the right hand side of Eq. \ref{Eq:TSVD2} represent the signal and noise components, respectively.
That is, the noise-reduced data $X_{svd}$ are obtained as
\begin{equation}
    X_{svd} =  U_r S_r V_r^{\mathsf{T}} = \sum_{i=1}^{r}  \bm{u}_i  \sigma_i  \bm{v}_i^{\mathsf{T}}. \label{Eq:TSVD3}
\end{equation}
The truncation rank $r$ is often selected based on the cumulative sum of singular values or prior information, such as the signal periodicity.

Note that the SVD is identical to the POD method; however, in the POD method, mean subtraction from the input dataset is essential.

\subsection{Noise reduction by the projected MSSA} \label{Sec:MSSA}
In this study, we examine the noise-reduction performance of the projected MSSA, which is composed of the MSSA and the low-dimensional representation using the SVD basis.

\subsubsection*{Step 1: Projecting data onto the subspace}
Using SVD, the input data $X$ can be projected onto the subspace spanned by the first $r$ singular vectors $U_r$:
\begin{equation}
    \tilde X = U_r^{\mathsf{T}} X = S_r V_r^{\mathsf{T}}. \label{Eq:TSVD4}
\end{equation}
The $\tilde X$ represents the low-dimensional representation of the input data and is a $r \times m$ matrix.
We denote the $i$-th row and element of $\tilde X$ as $\bm {\tilde x}_i$ and ${\tilde x}_{i,j}$, respectively.
\begin{equation}
    \tilde X = 
		\begin{bmatrix}
		   \bm{\tilde x}_{1} \\
		   \vdots \\
		   \bm{\tilde x}_{r}
		\end{bmatrix}
		=
		\begin{bmatrix}
		   \tilde{x}_{1,1} & \cdots  & \tilde{x}_{1,m}\\
		   \vdots  & \ddots & \vdots\\
		   \tilde{x}_{r,1} & \cdots  & \tilde{x}_{r,m}
		\end{bmatrix}.\label{Xtilde}
\end{equation}
As shown in Eqs. \ref{Eq:TSVD3} and \ref{Eq:TSVD4}, $\bm {\tilde x}_i$ (denoted as the temporal coefficient hereinafter) represents the temporal change in the expansion coefficients for $\bm u_i$. For example, if $r = r_X$, the $j$-th snapshot can be expressed using $\tilde x_{i,j}$ as
\begin{equation}
    {\bm x}(j) = \tilde{x}_{1,j} \bm{u_1} + \tilde{x}_{2,j} \bm{u}_2 + \cdots + \tilde{x}_{r,j} \bm{u}_r. \label{eq:reconstruction}
\end{equation}

Thereafter, the noise reduction is implemented by applying MSSA to $\tilde X$.
In other words, this algorithm applies MSSA to the SVD-based denoised data via projected low-dimensional representation.
The MSSA consists of the following steps.

\subsubsection*{Step 2: Embedding}
The SSA reduces noise by extracting smooth trajectories in a state space.
The state-space reconstruction is performed by time-delay embedding as follows.
We define the trajectory matrix $\tilde T_i$ of $\bm{\tilde x}_i$ as
\begin{equation}\label{eq:trajectory_matrix}
		{\tilde{T_i}}=
		\begin{bmatrix}
		   \tilde{x}_{i,1} & \tilde{x}_{i,2} & \cdots  & \tilde{x}_{i,K}\\
		   \tilde{x}_{i,2} & \tilde{x}_{i,3} & \cdots  & \tilde{x}_{i,K+1}\\
		   \vdots  & \vdots  & \ddots & \vdots\\
		   \tilde{x}_{i,L} & \tilde{x}_{i,L+1}  &\cdots  & \tilde{x}_{i,m}
		\end{bmatrix}.
\end{equation} 
Here, $L$ and $K$ represent the window length and the window width, respectively, and $K = m-L+1$.
Each column of $\tilde T_i$ is a partial time-series with the length of $L$ extracted from the time series $\bm {\tilde x}_i$ and represents a state-space vector converted from $\bm {\tilde x}_i$ by time-delay embedding.
$\tilde T_i$ has the same elements along the antidiagonals.
The multivariate trajectory matrix ${\tilde{T}}$ is constructed by stacking each trajectory matrix $\tilde T_i$:
		\begin{equation}
		{\tilde{T}}=
		\begin{bmatrix}
		   \tilde{T}_1 \\
		   \vdots \\
		   \tilde{T}_r 
		\end{bmatrix}.
		\end{equation}
$\tilde T$ is a $rL \times K$ matrix.

\subsubsection*{Step 3: SVD and grouping}
In general SSA procedures, the trajectory matrix undergoes SVD, and the decomposed components are grouped by some criteria.
Here, the grouping is performed following the same procedure employed for noise reduction using the truncated SVD method (Section \ref{Sec:SVD}).
That is, we decompose the trajectory matrix $\tilde T$ by SVD as ${\tilde{T}}={P}{D}{Q}^\mathsf{T}$, and assume that the noise components can be separated as the components corresponding to small singular values. 
Then, the noise-reduced trajectory can be extracted as
\begin{equation}\label{eq:SVD_POD}
		  \hat{T}=
		\begin{bmatrix}
		   \hat{T}_1 \\
		   \vdots \\
		   \hat{T}_r 
		\end{bmatrix}
		=
		  {P}_{r_{mssa}}{D}_{r_{mssa}}{Q}_{r_{mssa}}^\mathsf{T}
\end{equation}
Here, we set the truncation rank $r_{mssa}$ to $r_{mssa}=r$.

\subsubsection*{Step 4: Diagonal averaging}
Since the trajectory matrix has elements corresponding to temporal coefficients of the same snapshot along the antidiagonals (Eq. \ref{eq:trajectory_matrix}), 
the noise-reduced trajectory matrix $\hat T_i$ can be transformed into the temporal coefficient $\bm{\hat x}_{i}$ by obtaining the average along the antidiagonals:
\begin{equation}
    {\hat{x}_{i,j}}=\sum_{(a,b)\in A_j}\frac{(\hat{T}_i)_{a,b}}{n(A_j)},
\end{equation}
and 
\begin{equation}
    \hat X = 
		\begin{bmatrix}
		   \bm{\hat x}_{1} \\
		   \vdots \\
		   \bm{\hat x}_{r}
		\end{bmatrix}
		=
		\begin{bmatrix}
		   \hat{x}_{1,1} & \cdots  & \hat{x}_{1,m}\\
		   \vdots  & \ddots & \vdots\\
		   \hat{x}_{r,1} & \cdots  & \hat{x}_{r,m}
		\end{bmatrix}\label{Xhat}.
\end{equation}
Here, $(\hat{T}_i)_{a,b}$ is the element in the $a$-th row and $b$-th column of $\hat T_i$.
$A_j$ represents a set of indices $(a, b)$, comprising the antidiagonals of $\hat T_i$, i.e., $A_j=\{(a,b):a+b-1=j, 1\leq a \leq L, 1\leq b \leq K\}$.
$n(A_j)$ is the number of elements in $A_j$.
		
\subsubsection*{Step 5: Reconstructing the noise-reduced data}
Finally, the noise-reduced time-series data $X_{mssa}$ can be obtained in a manner similar to Eq. \ref{eq:reconstruction} as
\begin{equation}
    X_{mssa} =  U_r \hat{X}.
\end{equation} 

\section{Numerical experiments using fluid-simulation data}
\subsection{Input data}
First, the noise-reduction performances of TSVD and PMSSA were evaluated for the pressure field of a K\'{a}rm\'{a}n vortex street behind a square cylinder obtained by spatial two-dimensional fluid simulation.
The Mach number of the flow field was 0.2, and the Reynolds number was 120, defined using the length $l$ of the side of the square cylinder and the velocity $U_\infty$ of the uniform flow as reference values.
For details on this fluid simulation, please refer to the previous study \citep{Ohmichi2022}.
The pressure-coefficient distribution of the wake region, i.e., $x/l = [0,~10]$ and $y/l = [-5,~5]$, sampled into equally spaced $101 \times 101$ grid points was used as the input dataset.
The center of the cylinder was located at the origin $(x/l, y/l) = (0, 0)$.
The number of snapshots was $m = 1000$, and the time interval between the snapshots was $\Delta t = 0.125 l/U_\infty$.
The shot noise could be modeled as random normal noise.
To investigate the noise-reduction performance, random normal noise with a standard deviation $\sigma$ was added to each snapshot as the observation noise.

\subsection{Results and discussion}
\begin{figure}[hbt!]
    \centering
    \includegraphics[width = .95\textwidth]{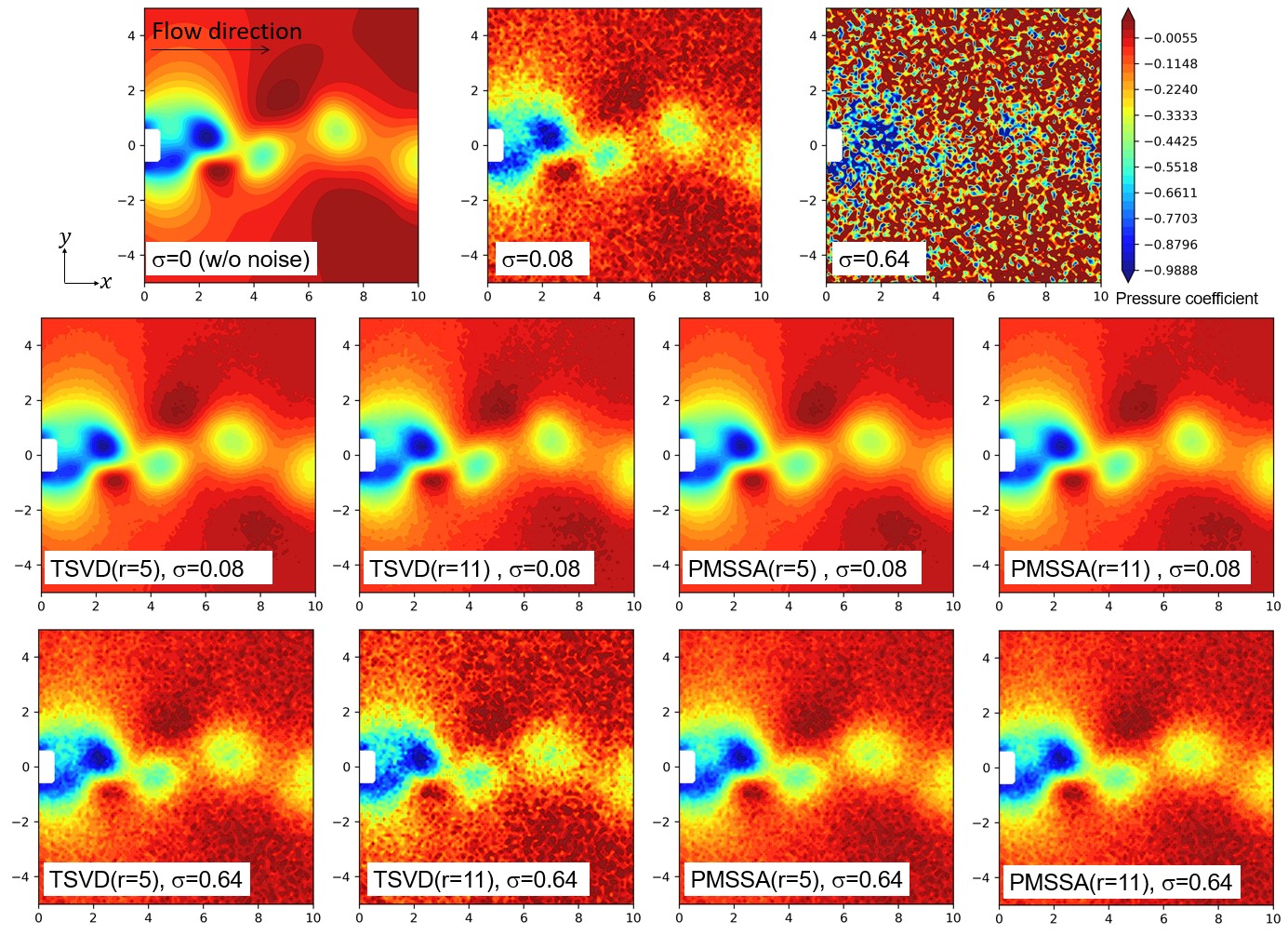}
    \caption{Input-pressure distribution images with different noise levels on the top row, and the noise-reduced images with TSVD and PMSSA on the middle ($\sigma = 0.08$) and bottom ($\sigma = 0.64$) rows.  }
    \label{fig:prism_images}
\end{figure}

Figure \ref{fig:prism_images} shows the noise-reduced images by TSVD and PMSSA.
The top row shows the input-pressure distribution with noise values of $\sigma = 0, ~0.08$, and 0.64.
The middle and bottom rows show the results of the noise reduction for the input data with $\sigma = 0.08$ and $0.64$, respectively.
The window length in PMSSA was set to $L = 90$, and the truncation ranks were set to $r = 5$ and 11.
Figure \ref{fig:prism_images} indicates that when the noise was small $\sigma = 0.08$, the differences between TSVD and PMSSA and the different $r$ results were visually small.
The result of PMSSA had a slightly smoother distribution at high noise level $\sigma = 0.64$ than that of TSVD.

To quantitatively evaluate the noise-reduction performance, we define the relative error as follows:
\begin{equation}
    Relative~error = \frac{{\left\lVert X_{rec} - X_{0} \right\rVert}_F}{{\left\lVert X_{0} \right\rVert}_F}
\end{equation}
$X_{rec}$ represents the data denoised by TSVD and PMSSA (i.e., $X_{svd}$ and $X_{mssa}$), and $X_0$ is the original data without noise ($\sigma = 0$).

\begin{figure}[hbt!]
    \centering
    \includegraphics[width = .4\textwidth]{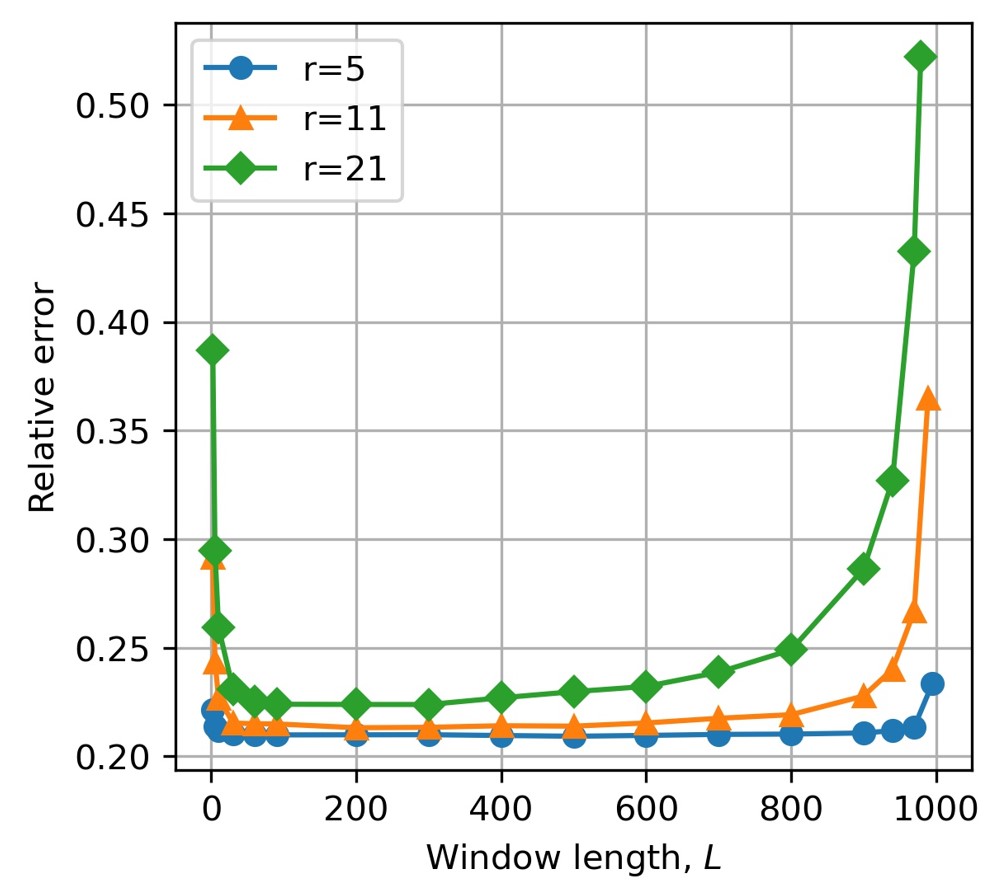}
    \caption{Effect of the window length $L$ on the relative error. $\sigma = 0.64.$}
    \label{fig:effect_of_L}
\end{figure}

The effect of the window length $L$ on the relative error in PMSSA is shown in Fig. \ref{fig:effect_of_L}.
The appropriate value of $L$ may depend on the nature of the signal and noise; however, \cite{Golyandina2010} reported that $L = m/2$ is a good choice in many cases for the basic SSA.
Figure \ref{fig:effect_of_L} shows that $L=m/2$ exhibited a small error here as well.
However, in the interval of $30~(\approx \sqrt{m})<L<500~(=m/2)$, the difference in error was relatively small.
The size of the trajectory matrix is $rL\times(m-L+1)$; thus, the smaller the $L$, the lower the computational cost.
An empirical guideline for achieving the appropriate $L$ involves choosing a value in the order of $\sqrt{m}$.
In this problem, we set $L = 90~(\approx3\sqrt{m})$.

\begin{figure}[hbt!]
    \centering
    \includegraphics[width=.4\textwidth]{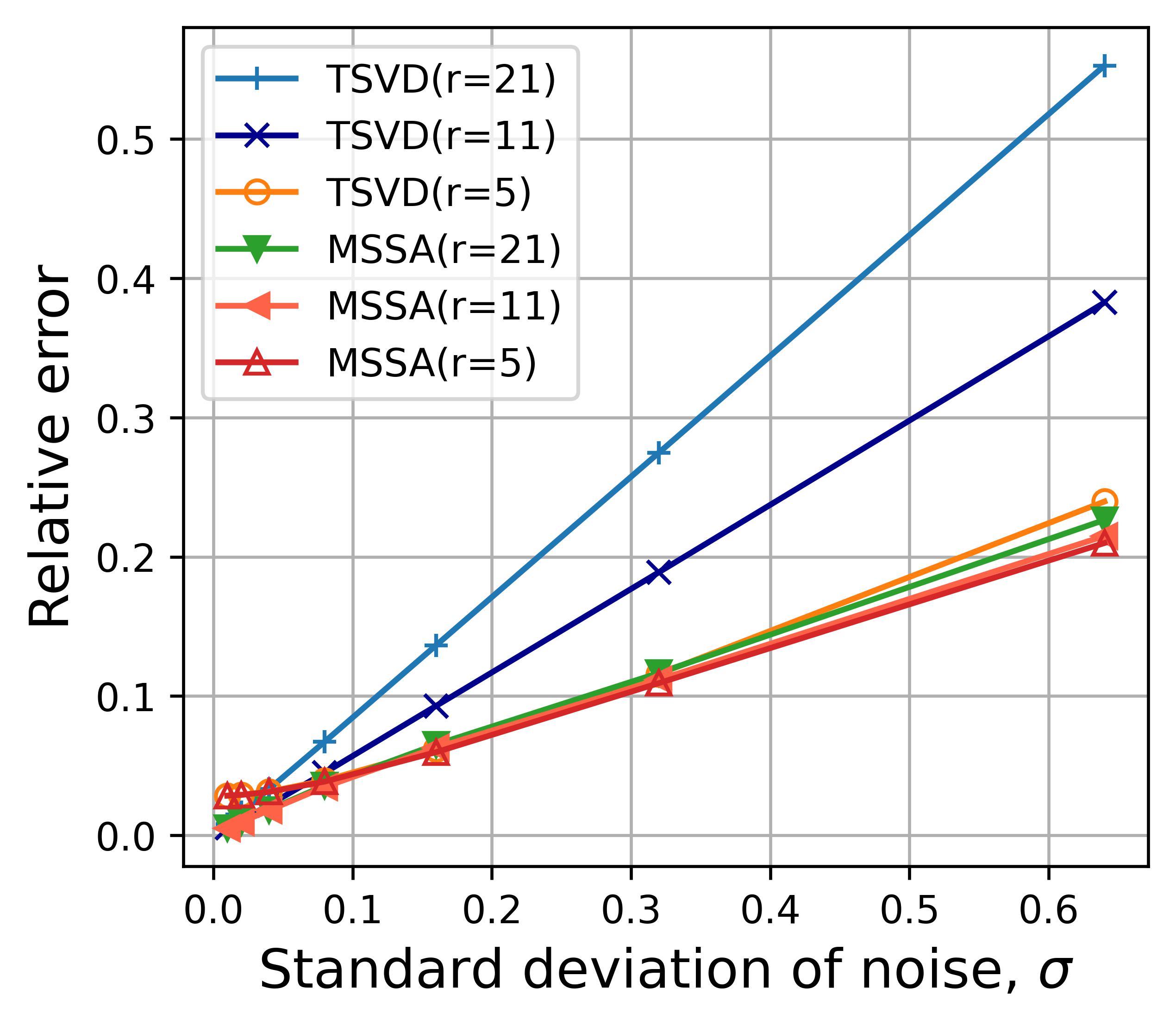}
    \caption{Effect of the truncation rank $r$ on the relative error in TSVD and PMSSA.}
    \label{fig:effect_of_sigma}
\end{figure}

Figure \ref{fig:effect_of_sigma} shows the change in error for different noise levels at each truncation rank.
For TSVD, the difference in the truncation rank had a significant impact on the error.
The error was the smallest at $r = 5$ at high noise levels.
At low noise levels, the errors were smaller at $r = 11$ and $r = 21$ than that of $r=5$ because even the signal component was truncated.
PMSSA is less sensitive to the truncation rank, and a larger truncation rank than that for TSVD can be applied even at high noise levels.
When $r = 11$ and 21, small errors were achieved at all noise levels.
At sufficiently low noise levels, the errors of TSVD and PMSSA were comparable.

\begin{figure}[hbt!]
    \centering
    \includegraphics[width=.95\textwidth]{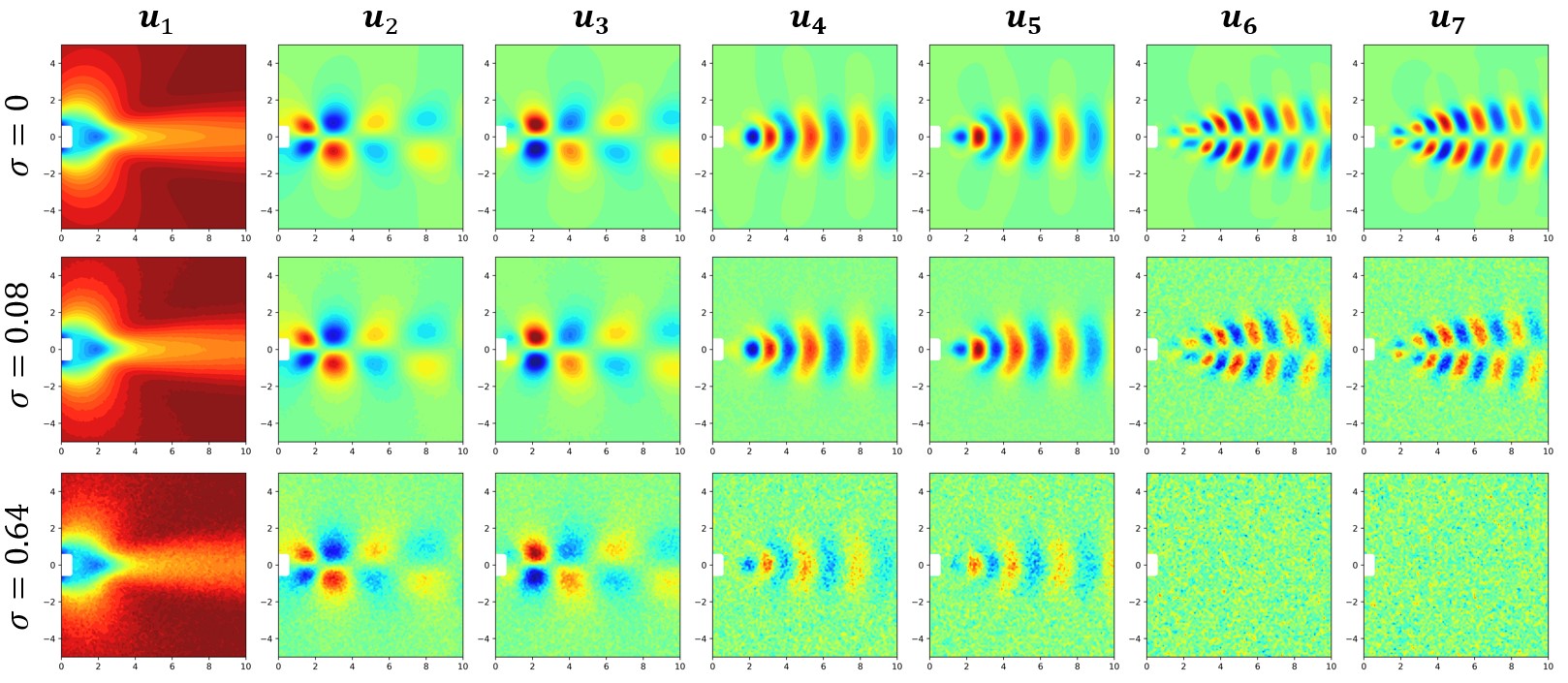}
    \caption{Spatial modes $\bm u_i$ extracted by SVD at different noise levels.}
    \label{fig:prism_mode}
\end{figure}

Figure \ref{fig:prism_mode} shows the spatial modes $\bm u_i$ extracted by SVD at each noise level.
The first mode $\bm u_1$ represents the mean field, while the second and third modes represent the pressure patterns indicating K\'{a}rm\'{a}n vortex shedding.
The fourth/fifth and sixth/seventh mode pairs represent the second and third harmonics of the K\'{a}rm\'{a}n vortex mode, respectively, while the sixth and seventh modes at $\sigma = 0.64$ did not exhibit coherent patterns because the noise level was extremely high.

\begin{figure}[hbt!]
    \centering
    \includegraphics[width=.85\textwidth]{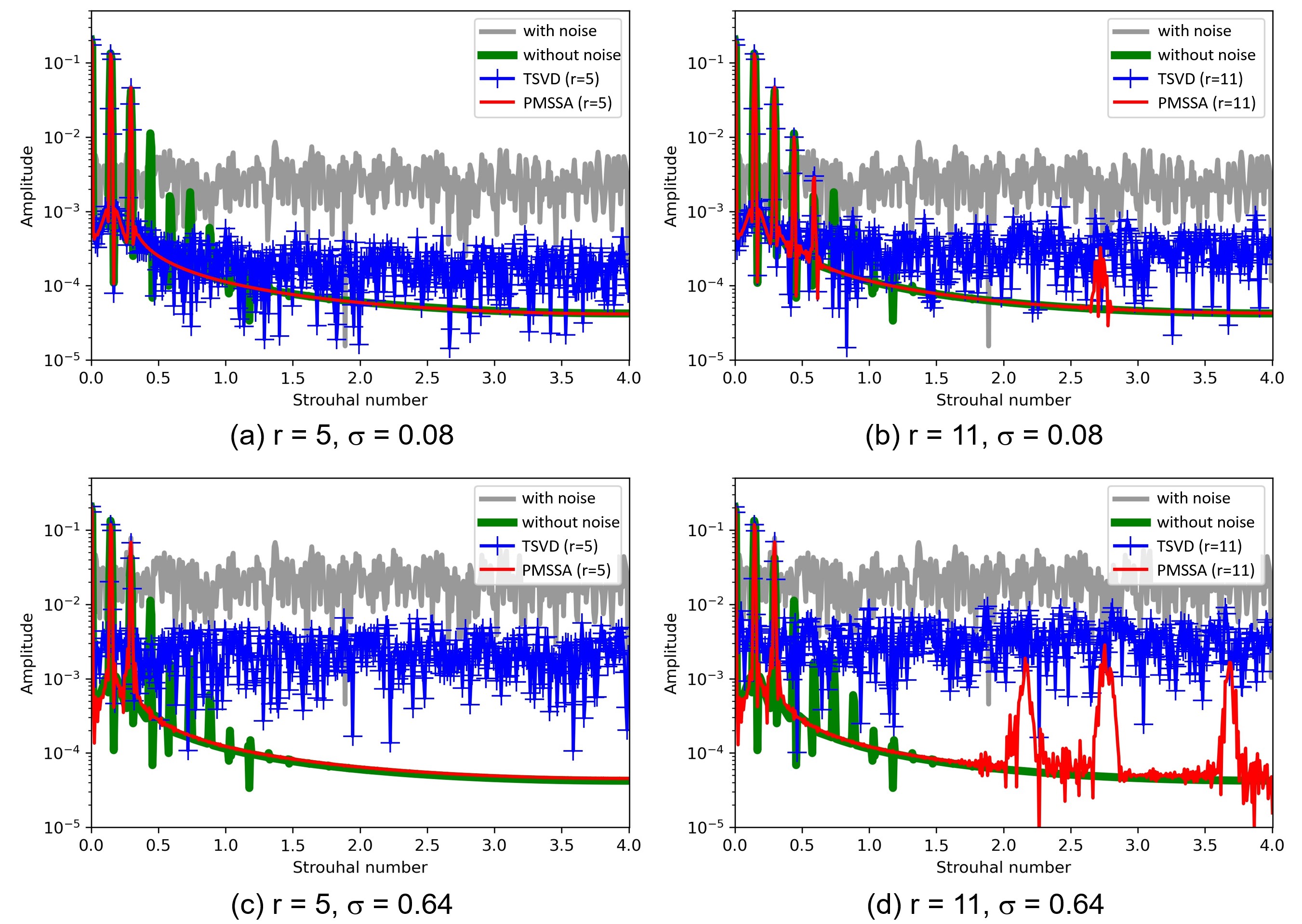}
    \caption{Comparison of the pressure signals at the point, ($x/l,~ y/l$) = (3.0, 0.5), behind the square cylinder in the frequency domain. }
    \label{fig:prism_frequency}
\end{figure}

\begin{figure}[hbt!]
    \centering
    \includegraphics[width=.7\textwidth]{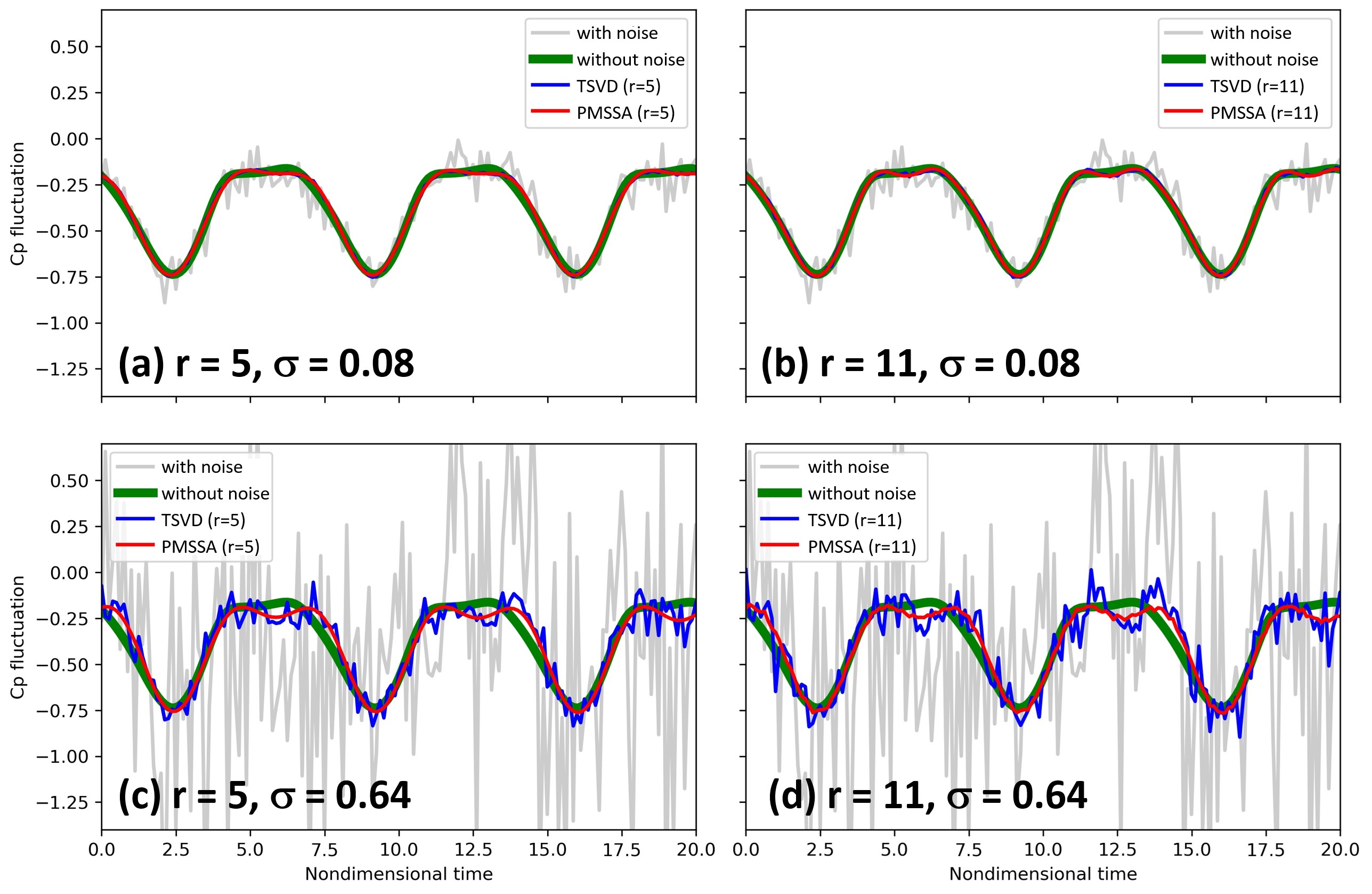}
    \caption{Comparison of the pressure signals at the point, ($x/l,~ y/l$) = (3.0, 0.5), behind the square cylinder in the time domain. }
    \label{fig:prism_time_domain}
\end{figure}

Next, we examined the reproducibility of the time-series change.
Figure \ref{fig:prism_frequency} shows the frequency distribution for the pressure at the point, ($x/l,~ y/l$) = (3.0, 0.5), behind the square cylinder.
For each truncation rank, the frequency distributions for different noise levels are compared.
Figure \ref{fig:prism_time_domain} shows the comparison in the time domain.
From these figures, we can see that PMSSA was superior to TSVD in removing the noise components under all conditions.
Figure \ref{fig:prism_frequency} shows that TSVD could extract periodic signals, although the noise remained at frequencies different from those of the periodic signal components.
Conversely, PMSSA reduced noise over a wide frequency range while retaining the periodic signal representing a K\'{a}rm\'{a}n vortex shedding phenomenon.
This noise-reduction performance of PMSSA was not significantly affected by the truncation rank whereas the remaining noise increased with the truncation rank for TSVD.
Figures \ref{fig:prism_frequency}a and \ref{fig:prism_frequency}b show that at low noise levels ($\sigma=0.08$) and with an overly small truncation rank ($r = 5$), even the signal components, which appeared as harmonic peaks at $r = 11$, were truncated.
At high noise levels, this harmonic peak was not captured even at $r = 11$ because the amplitude of the harmonic signal was smaller than the noise floor.
This is a limitation of SVD-based methods: since both PMSSA and TSVD reconstruct the signal using SVD modes, if the SVD does not capture a target phenomenon, these methods cannot reproduce the phenomenon.

\begin{figure}[hbt!]
    \centering
    \includegraphics[width=.99\textwidth]{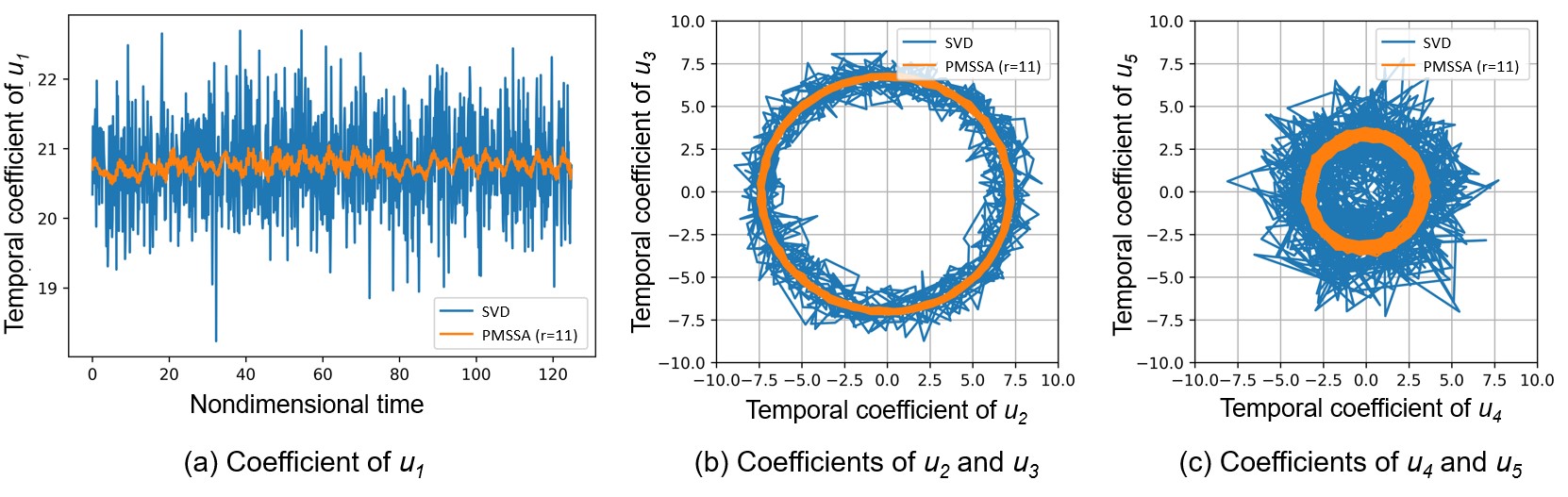}
    \caption{Comparison of the temporal coefficients $\bm{\tilde x}_i$ and $\bm{\hat x}_i$ of TSVD and PMSSA noise-reduced data for fluid-simulation data analyses. $\sigma = 0.64$.}
    \label{fig:prism_temporal_coefficients}
\end{figure}

Finally, we compared the temporal coefficients $\bm{\tilde x}_i$ and $\bm{\hat x}_i$ (Eqs. \ref{Xtilde} and \ref{Xhat}).
As explained in Section \ref{Sec:MSSA}, these are the results of the noise reduction by TSVD and PMSSA, expressed as the expansion coefficient for $\bm u_i$.
Figure \ref{fig:prism_temporal_coefficients}a shows the time history of the coefficient for $\bm u_1$, which is the mean mode.
Evidently, the coefficient for the mean mode fluctuated significantly in the TSVD, while the fluctuation was suppressed in the PMSSA.
Figures \ref{fig:prism_temporal_coefficients}b and \ref{fig:prism_temporal_coefficients}c show the phase plots for the coefficients of $\bm u_2,~\bm u_3$ and $\bm u_4,~\bm u_5$, respectively.
These pairs represent the K\'{a}rm\'{a}n vortex and its second-harmonic oscillation, respectively.
The figures clearly show that while PMSSA was able to extract smooth trajectories representing the relationship between the modes, the TSVD result retained a significant amount of noise in the trajectories.
These results indicate that PMSSA was superior in extracting the smooth trajectories of physical signals from noise-contaminated signals.

\section{Demonstration using PSP data}
\subsection{Input data}
\begin{figure}[hbt!]
    \centering
    \includegraphics[width=.95\textwidth]{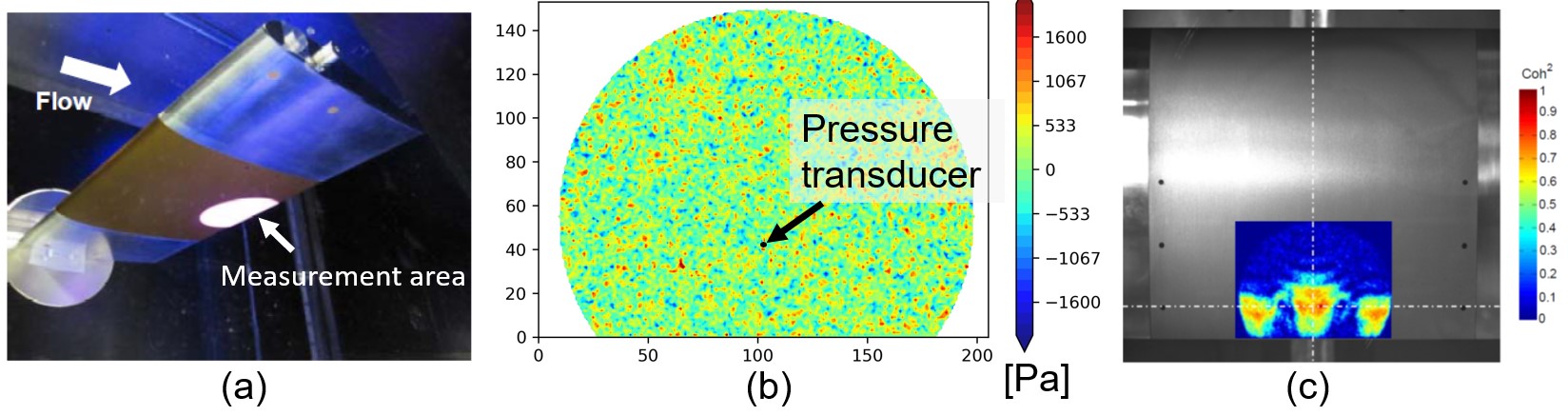}
    \caption{(a) NACA0012 airfoil model and measurement area for the PSP measurement. (b) Instantaneous pressure-distribution image contaminated by shot noise. (c) Coherence map of the 918 Hz signal taken by \cite{Nakakita2013}. }
    \label{fig:naca0012}
\end{figure}

As a demonstration using practical experimental data, we applied the PMSSA and TSVD to PSP data containing large noise.
The analyzed data were the time-series measurement data of the wing surface pressure distribution for the tone TE noise phenomenon on the NACA0012 airfoil \citep{Nakakita2011, Nakakita2013}.
The pressure distribution was measured using anodized aluminum PSP, which is one of the fast PSPs.
The low-turbulence calibration wind tunnel at JAXA Chofu Aerospace Center was employed for this experiment.
Figure \ref{fig:naca0012} shows the experimental model and the measured pressure-distribution image.
The main flow velocity was 28.0 m/s, the angle of attack was $-1.5^\circ$, and the pressure measurement area was a circular region on the pressure side of the wing (indicated by an arrow in Fig. \ref{fig:naca0012}a).
The sampling rate of the pressure-distribution image was 10 kHz, and the signals below 300 Hz were removed by a frequency filter to eliminate the dominant low-frequency noise, which was not random noise.
A $2\times2$ spatial averaging was performed, and the number of pixels after the spatial averaging was $d = 23917$.
The number of snapshots was $m = 10000$.
Figure \ref{fig:naca0012}b shows the pressure-distribution image after applying the frequency filter and spatial averaging.
Since the shot noise was large compared to the physical signal, coherent patterns could not be observed visually.
However, \cite{Nakakita2011, Nakakita2013} succeeded in extracting a coherent pressure pattern at approximately 918 Hz relevant to the tone TE noise by cross-spectrum analysis from this data (shown in Fig. \ref{fig:naca0012}c).
In the present analysis, we apply PMSSA to this data to demonstrate its effectiveness in reducing shot noise.
For details on this experiment, please refer to \cite{Nakakita2011, Nakakita2013}.

\subsection{Results and discussion}
\begin{figure}[hbt!]
    \centering
    \includegraphics[width = .95\textwidth]{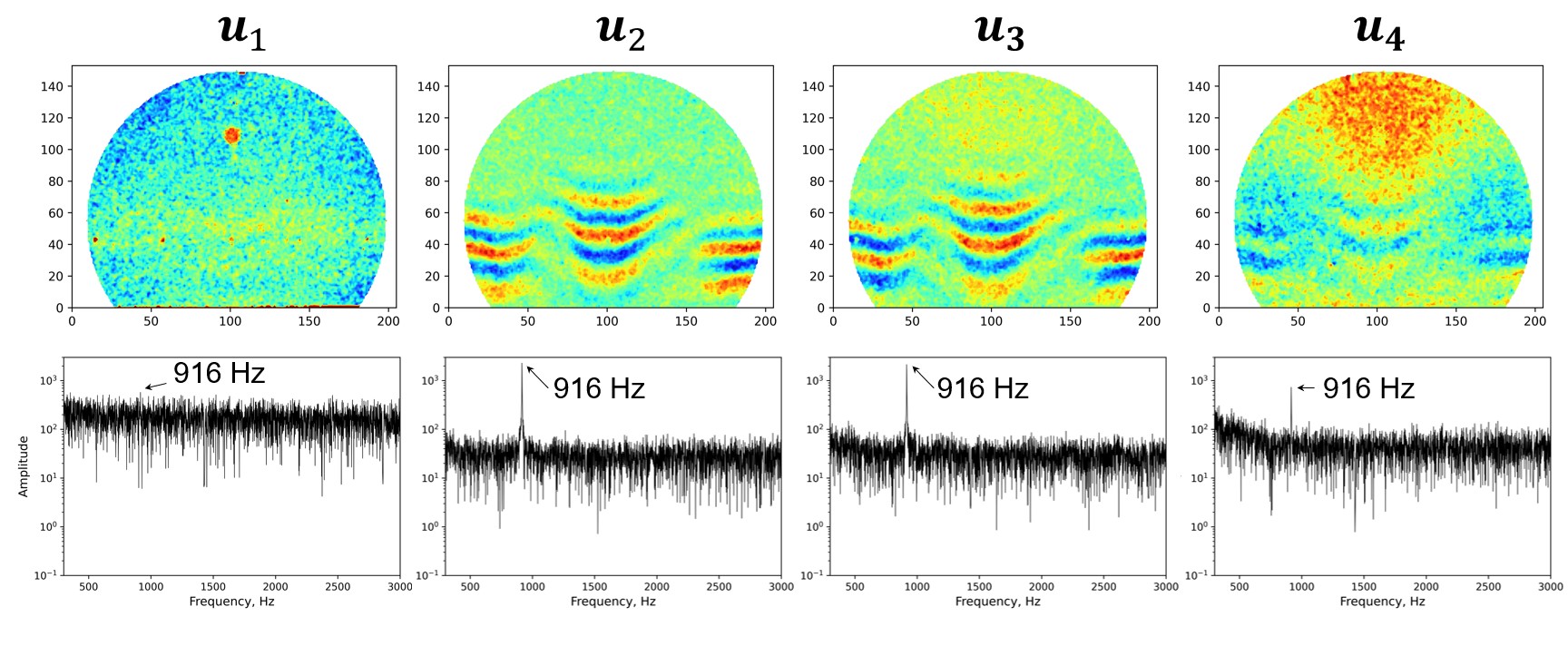}
    \caption{Spatial modes $\bm u_i$ extracted by SVD (upper row), and the frequency distributions of their temporal coefficients $\bm{\tilde x}_i$ (lower row). }
    \label{fig:svd_fft}
\end{figure}

First, we consider the results of the mode decomposition by SVD.
Figure \ref{fig:svd_fft} shows the spatial mode $\bm u_i$ obtained by SVD on the upper row and the frequency distribution of the temporal coefficient $\bm{\tilde x}_i$ corresponding to each mode on the lower row.
The most dominant mode, $\bm u_1$, had a noisy spatial distribution with a slightly coherent structure.
The frequency with the largest amplitude was 916 Hz, which corresponds to the fundamental frequency of the TE noise; however, the amplitude was similar in all frequency ranges, indicating that this mode behaved almost randomly in time.
Conversely, the second and third modes ($\bm u_2$ and $\bm u_3$) exhibited a clear wavy pattern, and the frequency distribution indicated that they displayed periodic oscillations at 916 Hz.
The fourth mode ($\bm u_4$) exhibited a similar wavy pattern with a peak frequency of 916 Hz, although different spatial patterns from the second and third modes were present.
In the frequency distribution of this mode, the amplitude of $O(100)$ Hz was slightly higher, and the different patterns shown in the spatial distribution could be considered to have $O(100)$ Hz fluctuations.
Although not shown in the figure, the fifth and subsequent modes did not have a clear peak frequency.

Based on the above discussion of the SVD mode, the truncation ranks for TSVD and PMSSA were set to $r = 4$ to capture the 916 Hz fluctuation pattern.
In PMSSA, the results at $r = 11$ were also compared to examine the effect of the truncation rank.
The window length was set to $L = 3\sqrt m = 300$.

\begin{figure}[hbt!]
    \centering
    \includegraphics[width=.95\textwidth]{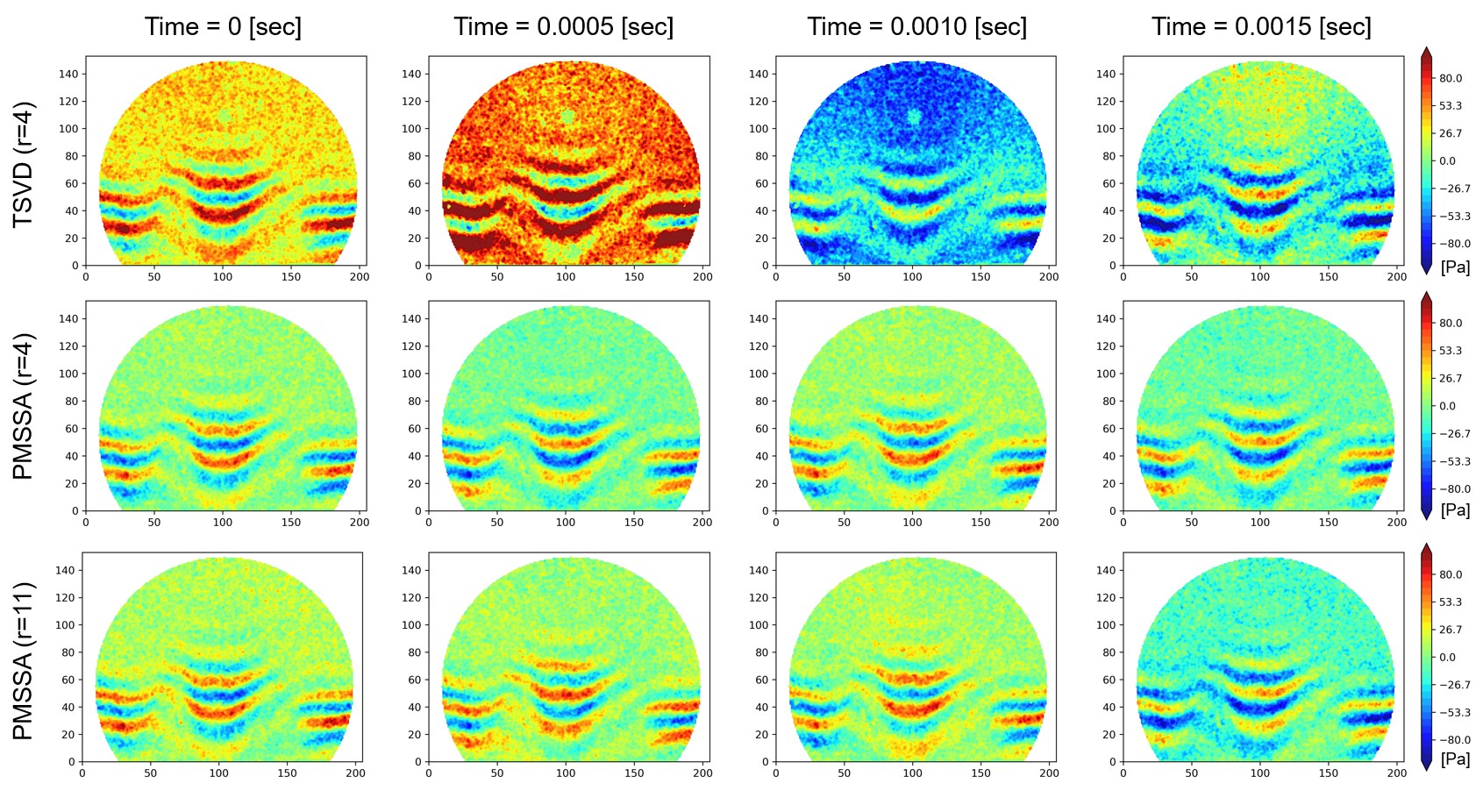}
    \caption{Reconstructed (i.e., denoised) pressure-distribution images obtained by TSVD ($r = 4$), PMSSA ($r = 4$), and PMSSA ($r = 11$). }
    \label{fig:denoised_psp}
\end{figure}

Figure \ref{fig:denoised_psp} shows the pressure-distribution image after the noise removal by the TSVD and PMSSA methods.
In the image before denoising (Fig. \ref{fig:naca0012}b), shot noise appeared in high levels, and coherent patterns were not observed.
Contrarily, after denoising by TSVD and PMSSA, in the image shown in Fig. \ref{fig:denoised_psp}, wavy patterns corresponding to the periodic fluctuation of 916 Hz appeared.
However, in the TSVD results, the pressure values of the entire region fluctuated significantly, which is attributed to the random-noise variation of $\bm u_1$ shown in Fig. \ref{fig:svd_fft}.
In PMSSA, only the temporal change in the wavy pattern was clearly extracted.
It was also observed that PMSSA was less sensitive to the truncation rank $r$, as it successfully removed the noise even when $r = 11$, which included an excessive number of modes to extract the 916 Hz oscillation.

\begin{figure}[hbt!]
    \centering
    \includegraphics[width=.95\textwidth]{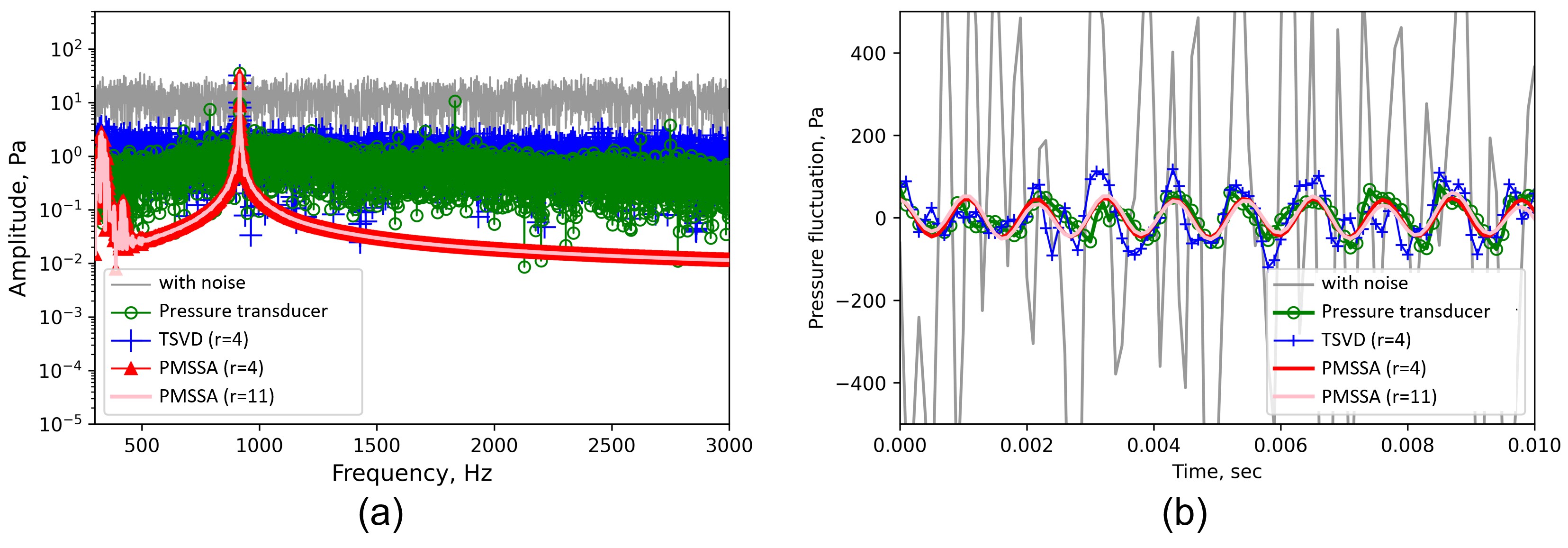}
    \caption{Comparisons of the denoised pressure signals and pressure transducer signal in the (a) frequency and (b) time domain.}
    \label{fig:comp_Kulite}
\end{figure}

In Fig. \ref{fig:comp_Kulite}, the comparison of the pressure fluctuations at the position indicated by the arrow in Fig. \ref{fig:naca0012}b is shown.
The result of the pressure measurement by the Kulite pressure transducer is also shown for comparison.
From the frequency distribution shown in Fig. \ref{fig:comp_Kulite}a, it is evident that the peak amplitudes at 916 Hz for both methods of TSVD and PMSSA were comparable to that by the pressure transducer.
However, for the TSVD, the noise levels in other frequency ranges were high.
PMSSA could significantly reduce the noise in both $r = 4$ and 11.
From the comparison of the pressure fluctuations in the time domain in Fig. \ref{fig:comp_Kulite}b, it was confirmed that the denoised signal with PMSSA exhibited less noise than that with TSVD, and a smooth temporal signal was extracted.
As shown in Fig. \ref{fig:comp_Kulite}a, the second-harmonic peak frequency of approximately 1840 Hz was captured by the pressure transducer, but not by TSVD and PMSSA.
This is because the amplitude of the pressure fluctuation at 1840 Hz was smaller than that of the shot noise generated in the PSP data, and the mode decomposition by SVD could not extract this phenomenon.

\begin{figure}[hbt!]
    \centering
    \includegraphics[width=.9\textwidth]{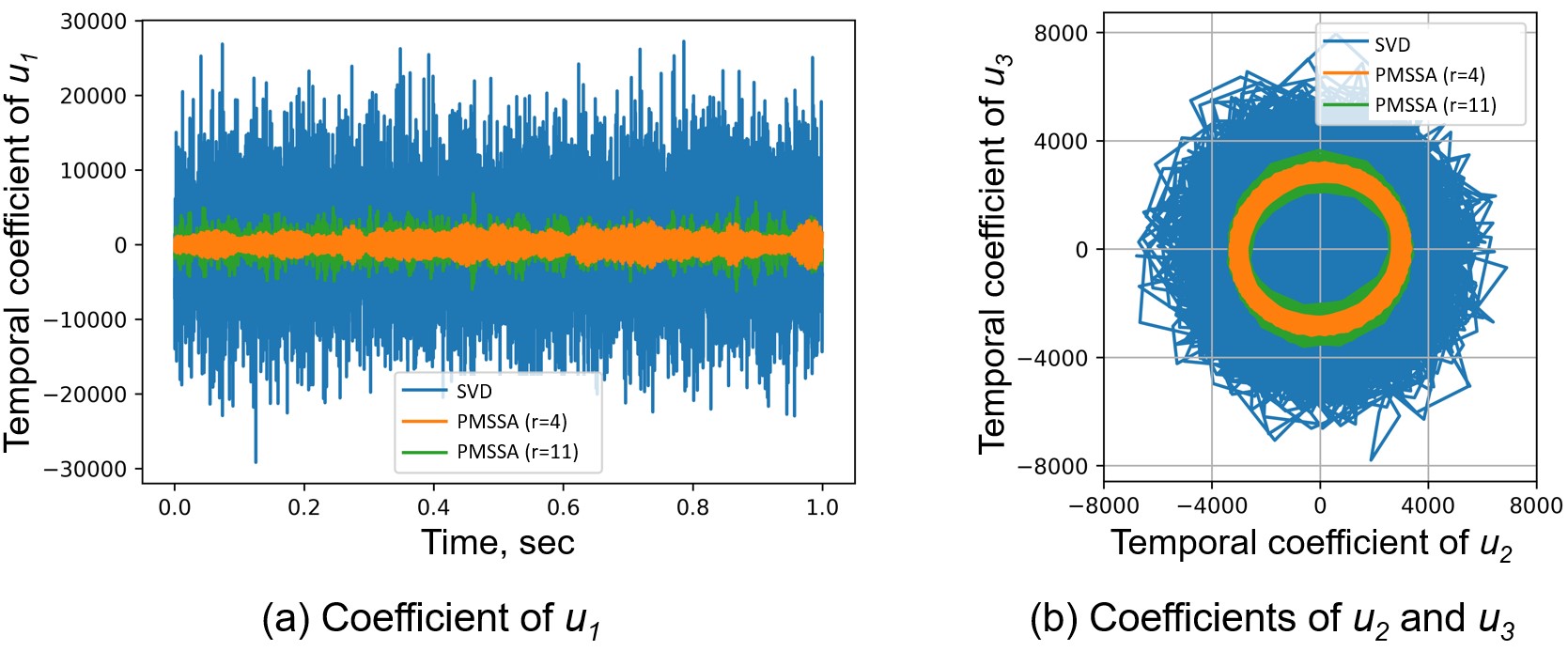}
    \caption{Comparison of the temporal coefficients $\bm{\tilde x}_i$ and $\bm{\hat x}_i$ of the TSVD and PMSSA noise-reduced data for the PSP data analysis.}
    \label{fig:tc_trajectory_psp}
\end{figure}

Figure \ref{fig:tc_trajectory_psp} shows a comparison of the temporal coefficients ($\bm{\tilde x}_i$ and $\bm{\hat x}_i$) obtained by the TSVD and PMSSA methods.
The time variation of the temporal coefficient for the first mode ($\bm u_1$) in Fig. \ref{fig:tc_trajectory_psp}a clearly shows that the PMSSA reduced the large random fluctuations remaining in the TSVD result.
In addition, in Fig. \ref{fig:tc_trajectory_psp}b, the phase plot represented by the coefficients for the second and third modes ($\bm u_2$ and $\bm u_3$) clearly shows that the PMSSA extracted a smoother trajectory than TSVD.
In other words, PMSSA can identify the modes behaving randomly and modes representing physical phenomena via modifying their trajectories.
The effective denoising is achieved by reconstructing PSP data based on the identified physically meaningful SVD modes and their modified trajectories.

\section{Conclusion}
A numerical method for reducing random noise from high-dimensional time-series data was investigated for denoising unsteady PSP data.
Unsteady PSP data are significantly affected by random noise generated by shot noise.
Here, we demonstrated the effectiveness of a noise-reduction method, i.e., the projected MSSA.
The projected MSSA comprises a combination of MSSA and low-dimensional data representation.
For the low-dimensional representation, we projected the data onto the subspace spanned by the SVD basis.
The pressure distributions for K\'{a}rm\'{a}n vortex shedding obtained by numerical fluid simulations and PSP data on the NACA0012 airfoil relevant to the tone TE noise were analyzed.
The results clearly showed several advantages of the projected MSSA.
The projected MSSA exhibited better noise-reduction performance than the conventional truncated SVD method in both analyses.
In addition, the projected MSSA was less sensitive to the truncation rank than the truncated SVD and exhibited better performance in a wide range of truncation rank values, since the projected MSSA can remove random variations from the SVD temporal coefficients.
Furthermore, the projected MSSA does not require prior information such as noise amplitude or pressure tap data as reference signals to identify noise components.
These advantages are achieved by extracting smooth trajectories of temporal coefficients in a state space by using time-delay embedding.
It should be noted, however, that since the projected MSSA uses the spatial modes from SVD for low-dimensional projection, signals not captured by SVD modes cannot be recovered.
The projected MSSA can be combined with mode decomposition methods other than SVD, and the development of mode decomposition methods with better detection limits is desired.

\backmatter

\section*{Acknowledgments}
The authors thank Hiroki Iwamoto for his technical support regarding the processing of the PSP data.

\section*{Declarations}
\subsection*{Ethical Approval}
Not applicable.

\subsection*{Competing interests}
The authors have no competing interests to disclose.

\subsection*{Authors' contributions}
YO contributed to conceptualization, data curation, funding acquisition, investigation, methodology, software, validation, visualization, writing -- original draft, and writing -- review and editing; KT contributed to investigation, software, validation, writing -- original draft, and writing -- review and editing; KN contributed to investigation, resources, and writing -- review and editing.

\subsection*{Funding}
This work was supported in part by the Japan Society for the Promotion of Science (JSPS) KAKENHI (grant no. 20K14958). 

\subsection*{Availability of data and materials}
The data that support the findings of this study are not publicly available. Data are however available from the authors upon reasonable request and with permission of Japan Aerospace Exploration Agency.


\bibliography{sn-bibliography}


\end{document}